\documentclass{article}
\usepackage{spconf,amsmath,graphicx}
\usepackage{xcolor}
\usepackage{amsfonts}
\usepackage{subcaption}
\usepackage{tikz}
\usepackage{tikz-cd}
\usetikzlibrary{positioning}
\usetikzlibrary{3d}
\usetikzlibrary{matrix}
\usetikzlibrary{decorations.text}
\usetikzlibrary{spy}
\usepackage{array}
\usepackage{algorithm}
\usepackage{algorithmic}
\usepackage{bbm}
\usepackage{hyperref}

\usepackage{mwe} 

\usepackage[permil]{overpic}
\usepackage{pict2e} 
\usepackage{listings}

\usepackage{tikz}
\usepackage{tikz-cd}

\usetikzlibrary{positioning}
\usetikzlibrary{3d}
\usetikzlibrary{matrix}
\usetikzlibrary{decorations.text}
\usetikzlibrary{spy}
\makeatletter
\tikzoption{canvas is plane}[]{\@setOxy#1}
\def\@setOxy O(#1,#2,#3)x(#4,#5,#6)y(#7,#8,#9)%
{\def\tikz@plane@origin{\pgfpointxyz{#1}{#2}{#3}}%
	\def\tikz@plane@x{\pgfpointxyz{#4+#1}{#5+#2}{#6+#3}}%
	\def\tikz@plane@y{\pgfpointxyz{#7+#1}{#8+#2}{#9+#3}}%
	\tikz@canvas@is@plane
}
\makeatother
\tikzoption{canvas is xy plane at z}[]{%
	\def\tikz@plane@origin{\pgfpointxyz{0}{0}{#1}}%
	\def\tikz@plane@x{\pgfpointxyz{1}{0}{#1}}%
	\def\tikz@plane@y{\pgfpointxyz{0}{1}{#1}}%
	\tikz@canvas@is@plane
}
\makeatother

\def\l{$\ell_0$}	

\definecolor{mygray}{gray}{0.5}

\usepackage{amsthm}
\newtheorem*{remark}{Remark}
\DeclareMathOperator*{\argmin}{arg\,min}

\title{Weighted-CEL0 sparse regularisation for molecule localisation in super-resolution microscopy with Poisson data}
 \name{Marta Lazzaretti$^1$, Luca Calatroni$^{2}$ and Claudio Estatico$^1$ \thanks{ML and LC acknowledges the support of UCA IDEX JEDI. The authors thank L. Blanc-F\'eraud for useful discussions and suggestions. ML carried out this work during her stage within the MORPHEME team (I3S, UCA).}}
 
 \address{
 $^{1}$ Universit\`a degli studi di Genova, DIMA, Italy\\
 $^{2}$ Université Côte d’Azur, CNRS, INRIA, I3S, France \\
}

\begin{document}
%
\maketitle
\begin{abstract}
We propose a continuous non-convex variational model for Single Molecule Localisation Microscopy (SMLM) super-resolution in order to overcome light diffraction barriers. Namely, we consider a variation of the Continuous Exact $\ell_0$ (CEL0) penalty recently introduced to relax the $\ell_2-\ell_0$ problem where a weighted-$\ell_2$ data fidelity is considered to model signal-dependent Poisson noise. For the numerical solution of the associated minimisation problem, we consider an iterative reweighted $\ell_1$ (IRL1) strategy for which we detail efficient parameter computation strategies. We report qualitative and quantitative molecule localisation results showing that the proposed weighted-CEL0 (wCEL0) model improves the results obtained by CEL0 and state-of-the art deep-learning approaches for the high-density SMLM ISBI 2013 dataset.
\end{abstract}

\section{Introduction}
\label{sec:intro}

Single Molecule Localisation Microscopy (SMLM) is a technique in fluorescence microscopy which has gained the attention of both the biological and the mathematical communities over the recent years as it allows to overcome the intrinsic limitations in optical resolution imposed by the light diffraction. SMLM techniques (among which we mention, PALM, STORM,\ldots) exploit an acquisition process which takes advantage of the absorption/emission properties of fluorescent molecules, which are sequentially activated and deactivated at random so as to limit the density of visible molecules in the sample. As a result, SMLM data consist of a stack of noisy and blurred images, whose individual frames represent sparse molecule samples which are easier to analyse and which can be re-combined at a final stage to obtain the desired super-resolved image. In terms of localisation precision, the quality of the result strongly depends on the density of the molecules activated at each frame and most of the existing models fail whenever such value is too high (see \cite{Sage2015} for a review).

In \cite{Gazagnes2017}, the authors considered a $\ell_2-\ell_0$-type continuous non-convex sparsity-promoting variational model for super-resolution of SMLM high-density data. Such model had been previously studied and thoroughly analysed in \cite{CEL02015} where \emph{exact} relaxation properties were shown to hold w.r.t.~to the original, NP-hard, $\ell_2-\ell_0$ model. The $\ell_2$ data term considered in these works was adapted to describe the presence of additive white Gaussian noise, although in \cite{Gazagnes2017} was shown to perform rather well also in the case of Poisson distributed data, which is a more realistic scenario in biological imaging.

In this work, we propose a sparsity-promoting weighted $\ell_2-\ell_0$-type model accounting more precisely for signal-dependent Poisson noise in SMLM data. Our modelling approximates the Kullback-Leibler data fidelity functional corresponding to the Poisson negative log-likelihood as a weighted $\ell_2$ data fidelity with local data intensity weights. Correspondingly, the CEL0-type associated penalty promotes sparsity depending both on the degradation model and local intensity information, favouring locally sparser solutions in correspondence of highly-corrupted pixels. To solve the corresponding composite non-convex optimisation problem, we consider an iterative-reweighted $\ell_1$ algorithm and provide some algorithmic details regarding its (challenging) implementation. We validate our model on the high-density SMLM ISBI 2013 dataset and compare the results with CEL0 \cite{Gazagnes2017} and Deep-STORM \cite{DeepStorm2018} solutions.

\section{WEIGHTED $\ell_2-\ell_0$ OPTIMISATION}
\label{sec:wCEL0}

\subsection{Inverse problem formulation}
\label{ssec:model}

Let $\mathbf{y}\in \mathbb{R}^{M^2}_+$ a vectorised  (in lexicographic order) $M\times M$ image acquired by means of a PALM/STORM technique and $\mathbf{x}\in\mathbb{R}^{N^2}_+$, with $N=LM$, the desired $N\times N$ image containing precise molecule localisations defined on a $L$-times finer grid with $L\in\mathbb{N}$. The acquisition process can be described as:
\begin{equation*}
    \mathbf{y}=\mathcal{P}(\mathbf{R_L H x}),
\end{equation*}
where, for $\mathbf{z\geq 0}$,  $\mathcal{P}(\mathbf{z})$ denotes the vector of realisations of Poisson random variables with parameters $z_i\geq 0$, $\mathbf{H}\in \mathbb{R}^{N^2 \times N^2}$ is the BCCB matrix corresponding to the the two-dimensional periodic convolution with a specific Gaussian Point Spread Function (PSF) $\mathbf{h}\in\mathbb{R}^{N^2}$ and $\mathbf{R_L}\in \mathbb{R}^{M^2\times N^2}$ is the down-sampling operator mapping the desired image from the fine grid to the coarser one. For shorthand notation, we further set $\mathbf{A}:= \mathbf{R_L H}\in \mathbb{R}^{M^2\times N^2}$. 

For $\lambda>0$, we consider the following non-convex sparsity-promoting model for computing a sparse approximation of $\mathbf{x}$ under the assumption that the data $\mathbf{y}$ is Poisson-distributed:
\begin{equation}  \label{eq:model}
    \hat{\mathbf{x}}\in\argmin_{\mathbf{x}\in\mathbb{R}^{N^2}}~ D_{KL}(\mathbf{Ax};\mathbf{y}) + \lambda \|\mathbf{x}\|_0 + i_{\geq 0}(\mathbf{x}),
\end{equation}
where $D_{KL}$ denotes the Kullback-Leibler fidelity term, which is derived via standard MAP estimation (see, e.g., \cite{Sawatzky2011}) and is defined for $\mathbf{v}\in \mathbb{R}^{M^2}_+$ as $D_{KL}(\mathbf{v};\mathbf{y}):=\sum_{i=1}^{M^2} d_{KL}(v_i;y_i)$ with $d_{KL}(v_i;y_i):= v_i-y_i\log(v_i)$. The indicator function  $i_{\geq 0}(\cdot)$ forces the desired solution $\mathbf{x}$ to be non-negative (since it represents molecule positions), while the regularisation term $\|\cdot\|_0$ denotes the $N^2$-dimensional $\ell_0$ pseudo-norm defined by:
\begin{equation*}
\|\mathbf{x}\|_{0} = \sum_{i=1}^{N^2} |x_i|_0~\quad\text{with }\quad|x_i|_0:=\begin{cases}
1 \ &\text{if }\ x_i\neq 0\\
0 \ &\text{if }\ x_i= 0.
\end{cases}
\end{equation*}
Dealing directly with the Kullback-Leibler functional $D_{KL}$ above makes the problem very challenging. To overcome such difficulties, several approximations of $D_{KL}$ can be considered, see \cite{Sawatzky2011}. We consider here a second-order Taylor approximation of $D_{KL}(\cdot;\mathbf{y})$ around $\mathbf{y}$ 
which leads to the following, symmetric weighted-$\ell_2$ data term:
\begin{align} \label{eq:fidelity}
    \|\mathbf{Ax}-\mathbf{y} \|^2_{\mathbf{W}} & := \langle \mathbf{Ax}-\mathbf{y}, \mathbf{W}(\mathbf{Ax}-\mathbf{y})\rangle \notag \\
    & = \sum_{i=1}^{M^2} 
    \frac{((\mathbf{A x})_i - y_i)^2}{y_i},  
\end{align}
where the weighted norm is defined in terms of the diagonal, positive definite matrix $\mathbf{W}=\text{diag}(\mathbf{ 1_{M^2}} ./\  \mathbf{y})\in\mathbb{R}^{M^2\times M^2}$ where $\mathbf{1_{M^2}}~./\ \mathbf{y}$  denotes the Hadamard element-wise division between the $M^2$-dimensional vector with all elements equal to one and $\mathbf{y}$.
This fidelity term can now be used in \eqref{eq:model} as an approximation of $D_{KL}$. It weights locally the least-square discrepancy by the inverse intensity of the given low-resolution data. This choice thus enforces a large/low fidelity whenever low/high signal (corresponding to locally low/high noise) is measured, respectively.

Hence, instead of \eqref{eq:model}, we consider the following simplified weighted $\ell_2-\ell_0$ problem:
\small{
\begin{equation}  \label{eq:problem_weight}
        \hat{\mathbf{x}}\in\argmin_{\mathbf{x}\in\mathbb{R}^{N^2}}~ G_{w\ell_0}:= \|\mathbf{Ax}-\mathbf{y} \|^2_{\mathbf{W}} + \lambda \|\mathbf{x}\|_0 + i_{\geq 0}(\mathbf{x}).
\end{equation}}
\normalsize
We remark that due to the presence of the $\ell_0$ pseudo-norm, problems in the form \eqref{eq:problem_weight} are known to be NP-hard. Several locally convergent methods can be alternatively used to solve these problems, such as, for instance, the Iterative Hard Thresholding (IHT) and branch and bounds algorithms, which, however, are often hard to be applied in the case of large-scale data. To overcome this issue, in recent years a new class of continuous non-convex penalties has been studied for the $\ell_2-\ell_0$ problem (see, e.g., \cite{CEL02015,Carlsson2019}), based also on the analytical properties of their local/global minimisers studied in \cite{Nikolova2013}. The general idea for this type of penalties is to consider \emph{continuous} non-convex relaxations of the $\ell_0$ pseudo-norm obtained by repeated application of Fenchel conjugation. The continuity of the relaxed functional allows for the use of standard optimisation algorithms, such as, for instance, the iterative reweighted $\ell_1$ (IRL1) algorithm \cite{Ochs2015}. We proceed similarly and consider a variation of the \emph{continuous exact} $\ell_0$ (CEL0) penalty introduced in \cite{CEL02015} for the $\ell_2-\ell_0$ problem which is better suited to deal with the data term \eqref{eq:fidelity}.

\subsection{A weighted-CEL0 (wCEL0) penalty}

To derive a continuous approximation of the  non-convex functional $G_{w \ell_0}$  in \eqref{eq:problem_weight}, we follow \cite{CEL02015} and compute its biconjugate functional by applying twice Fenchel conjugation. 
Similarly as in \cite{CEL02015}, the computations can be first performed in a one-dimensional setting and then extended to the multi-dimensional for general operators $\mathbf{A}$ via some technical considerations (see \cite{TesiMarta2020} for the details).

We consider the following continuous relaxation of $G_{w \ell_0}$:\begin{equation}  \label{eq:GwCEL0}
G_{\texttt{wCEL0}}(x):=\frac{1}{2}||\mathbf{A x}- \mathbf{y}||_{\mathbf{W}}^2+\Phi_{\texttt{wCEL0}}(\mathbf{x};\lambda)+ i_{\geq 0}(\mathbf{x}),    
\end{equation}
where, for $\lambda>0$, $\Phi_{\texttt{wCEL0}}(\cdot;\lambda)$ denotes the non-convex non-smooth continuous penalty defined by:
\small{
\begin{equation*}
\Phi_{\texttt{wCEL0}}(\mathbf{x};\lambda):=\sum_{i=1}^{N^2}\lambda -\frac{\|\mathbf{a}_i\|_{\mathbf{W}}^2}{2}\left(|x_i|-\frac{\sqrt{2\lambda}}{\|\mathbf{a}_i\|_{\mathbf{W}}}\right)^2 \mathbbm 1_{\{|x_i|<\frac{\sqrt{2\lambda}}{\|\mathbf{a}_i\|_{\mathbf{W}}} \}},
\end{equation*}}
\normalsize
and $\mathbf{a}_i = (a_{j,i})_j\in \mathbb{R}^{M^2}$ denotes the $i$
-th column of the operator $\mathbf{A}$. Here,  the computation of the weighted norm $\|\mathbf{a}_i\|^2_{\mathbf{W}}$ contained in the expression of the penalty term $\Phi_{\texttt{wCEL0}}$ encodes the dependence on the data $\mathbf{y}$ since, by definition:
\begin{equation} \label{eq:w_norm_col}
\|\mathbf{a}_i\|_{\mathbf{W}}^2 = \sum_{j=1}^{M^2} \frac{a_{j,i}^2}{y_j}.
\end{equation}
One can prove exact continuous relaxation properties which guarantee that the global minima of $G_{w\ell_0}$ are also global minima of $G_{\texttt{wCEL0}}$ and that  $G_{\texttt{wCEL0}}$ eliminates some local minimisers of $G_{w\ell_0}$. We further address the reader to \cite{Carlsson2019} where such properties are studied in general scenarios.

\begin{remark}[Comparison with CEL0]
Compared to the $\Phi_{\texttt{CEL0}}$ penalty considered in \cite{CEL02015,Gazagnes2017} for the standard $\ell_2-\ell_0$ problem, the new penalty $\Phi_{\texttt{wCEL0}}$ presents an explicit dependence  on both the model (i.e. the columns of the operator $\mathbf{A}$, as for CEL0) and the data $\mathbf{y}$. This reflects the intrinsic signal-dependence encoded into the considered Poisson modelling and, numerically, reflects into the introduction of a threshold which is different for each component $x_i$ of the solution (as it involves the $i$-th column of $\mathbf{A}$) and adapts to any data $\mathbf{y}$.
\end{remark}

\section{ALGORITHMIC IMPLEMENTATION}
\label{sec:Alg}


We detail here the computation of the model and algorithmic parameters required to minimise the functional $G_{\texttt{wCEL0}}$ of \eqref{eq:GwCEL0}. We follow \cite{Gazagnes2017} and consider the IRL1 algorithm whose pseudocode is reported in Algorithm \ref{alg:wCEL0}.

\smallskip

\textbf{Weighted column norms computation.} The computation of the $N^2$ weighted norms \eqref{eq:w_norm_col} $\|\mathbf{a}_i\|_{\mathbf{W}}$ of the $M^2$-dimensional columns of the operator $\mathbf{A}={\mathbf{R_L H}}$ is required for the computation of the penalty term $\Phi_{\texttt{wCEL0}}$. To do so, we proceed as follows.
Since the operator $\mathbf{R_L}$computes down-sampling via the sum of $L \times L$ neighbourhood pixel values, it can be viewed as a restriction of the two-dimensional periodic convolution operator with kernel  $\mathbf{k_L} \in \mathbb{R}^{N\times N}$ defined as
$$ (\mathbf{k_L} )_{i,j}= 
\begin{cases}
1, \qquad i,j \in \{ \frac{ML}{2}-\frac{L}{2}+1, \cdots, \frac{ML}{2}+\frac{L}{2} \}\\
0, \qquad \text{otherwise,}
\end{cases}
$$ 
where $L$ here is assumed even for simplicity. Indeed, by denoting by $\mathbf{K_L} \in \mathbb{R}^{N^2 \times N^2}$ the BCCB matrix corresponding to the kernel $\mathbf{k_L}$, we can compute any matrix-vector product with $\mathbf{A}\in\mathbb R^{M^2\times N^2}$ by means of horizontal and vertical $L$-equispaced selections of the result of the matrix-vector product with the BCCB matrix $\mathbf{A_E}=\mathbf{K_L H} \in \mathbb{R}^{N^2 \times N^2}$. In this way, although $\mathbf{A_E}$ is larger than the original $\mathbf{A}\in\mathbb{R}^{M^2\times N^2}$, it is fully BCCB, so that its usage only involves the two-dimensional FFTs of the kernels $\mathbf{k_L}$ and $\mathbf{h}$ and a $ O(N^2 \log N)$ numerical complexity. 
The enlarged structured matrix $\mathbf{A_E}$ allows us to compute the weighted norms $\|\mathbf{a}_i\|_{\mathbf{W}}$. Indeed, let us first insert the acquired $M \times M$ matrix image $\mathbf{Y}$ into the $N \times N$ matrix $\tilde{\mathbf{Y}}$ as follows
\small{
\begin{equation*}
    (\tilde{\mathbf{Y}})_{i,j}=
        \begin{cases}
        ({\mathbf{Y}})_{p+1,q+1} & \; \text{ if } \, i=1+Lp, \,j=1+Lq,\\
             &\;\;\;\;\;\;\;\;\;\;\;\;\; \text{ for } \, p,q=0,\cdots,M-1 \\
        0 &\text{ otherwise.}
        \end{cases}
 \end{equation*}}
\normalsize
Denoting now by $\mathbf{\tilde{y}} \in \mathbb{R}^{N^2}$ the vectorisation of $\mathbf{\tilde{Y}}$ and by $^{ \cdot 2}$ element-wise matrix square, we have that the matrix-vector product  $\mathbf{v}=\mathbf{A_E}^{\cdot 2} (\mathbf{1_{M^2} ./\ \tilde{y}})$ gives the vector $\mathbf{v} \in \mathbb{R}^{N^2}$, whose $i$-th component is just the weighted sum of the square of the elements of the $i$-th column of $\mathbf{A}$, since the corresponding weights are the values $\mathbf{1_{M^2}} ./\ \mathbf{\tilde{y}}$. These $N^2$ quantities just correspond to $\|\mathbf{a}_i\|_{\mathbf{W}}^2$, for $i=1,\cdots,N^2$. We stress that while the penalty term in \cite{Gazagnes2017} does not involve the computation of weighted norms, the described representation of the down-sampling operation as restricted convolution is crucial here for the actual (fast) computation of \eqref{eq:w_norm_col}.

\smallskip

\textbf{Backtracking.}  As efficient solver for the inner weighted-$\ell_1$ problems of the IRL1 algorithm, we use a Generalised FISTA (GFISTA) algorithm with adaptive backtracking of the Lipschitz constant $\mathcal{L}=\| \mathbf{A^T W A}\|\leq \|\mathbf{A}\|^2\|\mathbf{W}\|$ of the gradient of the data term \eqref{eq:fidelity}, where $\|\cdot\|$ denotes the operator norm, see \cite{GFISTA2019}. As it is well-known for forward-backward algorithms, an accurate estimate of the Lipschitz constant $\mathcal{L}$ is required to ensure good convergence properties.
However, due to the sub-multiplicative property of $\|\cdot\|$,  an estimate of the type $\|\mathbf{A}\|^2 \|\mathbf{W}\| = L^2 \max(\mathcal{F}(\mathbf{h}))^2 \min(\mathbf{Y})^{-1}$ 
with $\mathcal{F}(\cdot)$ being the 2D FFT,  tends, typically, to significantly overestimate $\mathcal{L}$ due to the possible small values very close to zero assumed by $\mathbf{y}$. This corresponds to consider extremely small step-sizes $\tau$, due to the convergence condition $\tau\in (0,1/\mathcal{L}]$, which may badly affect convergence speed. The use of a backtracking strategy providing at each iteration of the IRL1 inner loop a good estimate of $\mathcal{L}$ avoids this drawback. 



\smallskip

\textbf{Parameters.} We initialise the IRL1 algorithm for both CEL0 and wCEL0 models by choosing $\mathbf{x^0}= \mathbf{A^T y}$ and assess convergence by a joint criterion based on the relative difference between consecutive iterates, the difference of their corresponding function values  and a maximum number of iterations for both the inner and the outer loop. Finally, we consider an heuristic tuning of the regularisation parameter $\lambda>0$ for both methods by averaging the parameters optimising the results for 8 randomly chosen temporal frames. 


\begin{algorithm}[t]
\label{alg:wCEL0_algo}
\algsetup{linenosize=\tiny}
\small{
\caption{Weighted CEL0 (wCEL0) via IRL1}
\label{alg:wCEL0}
\begin{algorithmic}
\REQUIRE $\mathbf{y}\in\mathbb{R}^{M^2}, \mathbf{x}^0\in\mathbb{R}^{N^2}, \lambda>0$
\REPEAT 
\STATE update weights \  $\mathbf{\omega}_i^{\mathbf{x^k}}\in \partial \Phi_{\texttt{wCEL0}}{(\mathbf{x^k};\lambda)}$
\STATE $\mathbf{x}^{k+1}=\argmin\limits_{\mathbf{x}\geq 0} \frac12 \| \mathbf{y} - \mathbf{A} \mathbf{x} \|_{\mathbf{W}}^2
+ \lambda \sum\limits_{i=1}^{N^2}\mathbf{\omega}_i^{\mathbf{x^k}}|x_i|$
\UNTIL convergence
\RETURN $\mathbf{x}$ 
\end{algorithmic}}
\end{algorithm}

\section{NUMERICAL RESULTS}
\label{sec:results}

\begin{figure}[h!]    
\centering
\begin{subfigure}[b]{0.15\textwidth}
\centering
\includegraphics[width=\textwidth]{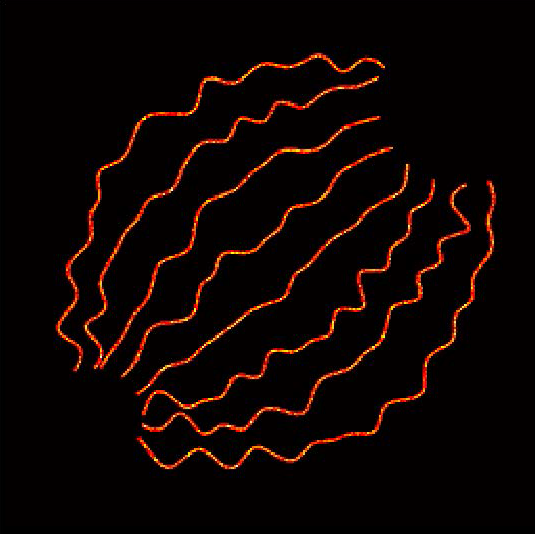}
\caption{GT}
\label{fig:gt}
\end{subfigure}
\begin{subfigure}[b]{0.15\textwidth}
\centering
\includegraphics[width=\textwidth]{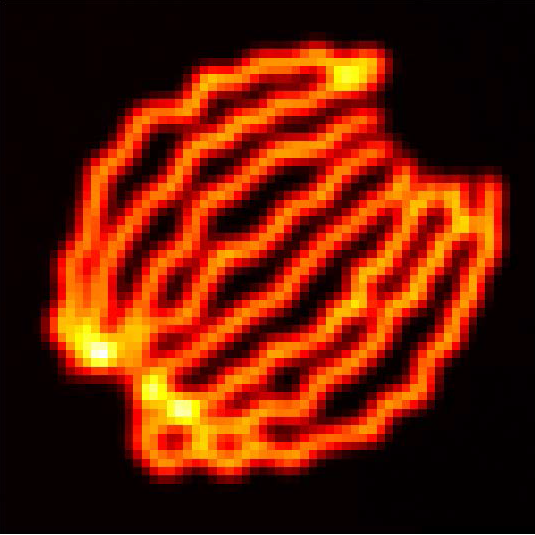}
\caption{$\overline{\mathbf{y}}$}
\label{fig:data}
\end{subfigure}
\begin{subfigure}[b]{0.15\textwidth}
\centering
\includegraphics[width=\textwidth]{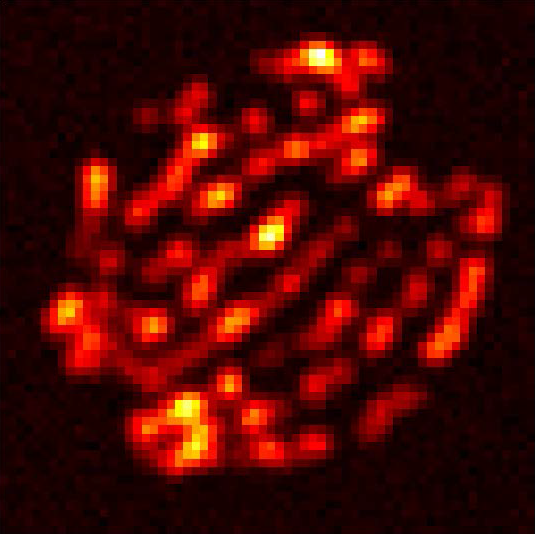}
\caption{Single frame}
\label{fig:frame}
\end{subfigure}
 \caption{ISBI SMLM 2013 dataset: (a) Ground Truth data, (b) sum of all acquisitions (x4), (c) single acquisition (x4).}
    \label{fig:zooms}
\end{figure}
\begin{figure}[h!]    
\centering
\begin{subfigure}[b]{0.15\textwidth}
\centering
\begin{overpic}[width=\textwidth]{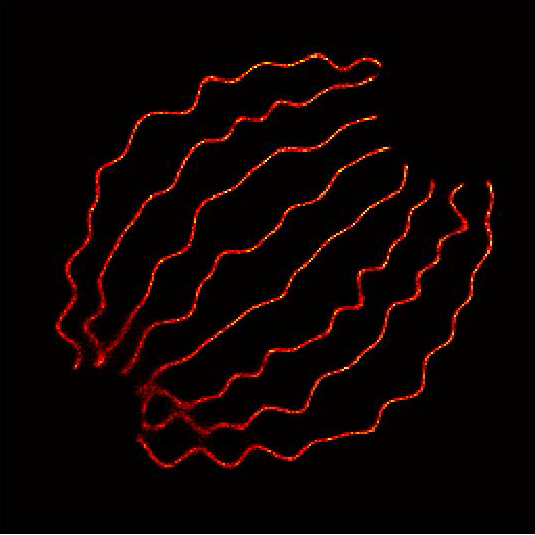}%
\put(230,110){\linethickness{0.25mm}\color{white}\polygon(0,0)(200,0)(200,200)(0,200)}%
\end{overpic}%
\caption{wCEL0}
\label{fig:wCEL0}
\end{subfigure}
\begin{subfigure}[b]{0.15\textwidth}
\centering
\begin{overpic}[width=\textwidth]{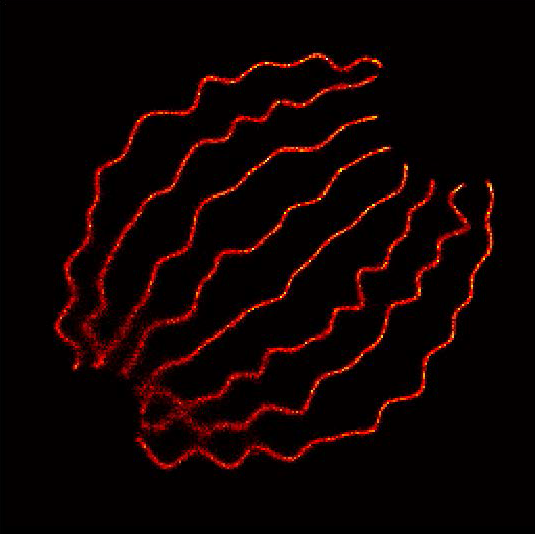}%
\put(230,110){\linethickness{0.25mm}\color{white}\polygon(0,0)(200,0)(200,200)(0,200)}%
\end{overpic}%
\caption{CEL0}
\label{fig:CEL0}
\end{subfigure}
\begin{subfigure}[b]{0.15\textwidth}
\centering
\begin{overpic}[width=\textwidth]{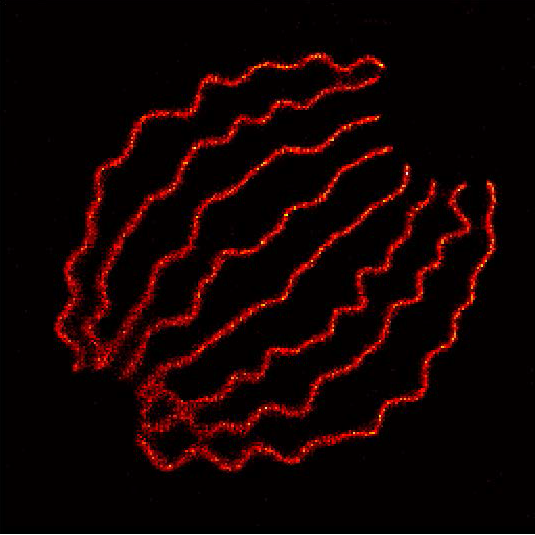}%
\put(230,110){\linethickness{0.25mm}\color{white}\polygon(0,0)(200,0)(200,200)(0,200)}%
\end{overpic}%
\caption{Deep-STORM}
\label{fig:DS}
\end{subfigure}\\
\begin{subfigure}[b]{0.15\textwidth}
\centering
\includegraphics[width=\textwidth]{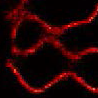}
\end{subfigure}
\begin{subfigure}[b]{0.15\textwidth}
\centering
\includegraphics[width=\textwidth]{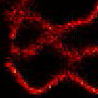}
\end{subfigure}
\begin{subfigure}[b]{0.15\textwidth}
\centering
\includegraphics[width=\textwidth]{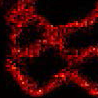}
\end{subfigure}
\caption{\textbf{First-row}: (a) wCEL0 result, (b) CEL0 result, (c) Deep-STORM result. \textbf{Second row}: close-up on a detail.}
\label{fig:close-ups}
\end{figure}

We test the proposed wCEL0 model on the high-density ISBI SMLM 2013 dataset where 217 fluorophores are activated on average at each time acquisition. The dataset is composed of $361$ images representing 8 tubes of $30$nm diameter. The size of each acquisition is $64\times 64$ pixels where each pixel is of size $100\times 100$nm$^2$.  We localise the molecules on a $256\times 256$ pixel image corresponding to a factor $L=4$, where the size of each pixel is thus $25\times 25$nm$^2$. The total number of molecules is $81049$. The Gaussian PSF has  $\text{FWHM}=258.2$nm. We report in Fig. \ref{fig:gt} the ground-truth image, in Fig. \ref{fig:data} the sum of all acquisitions, and in Fig. \ref{fig:frame} an example of a single frame from the dataset. We report in Figure \ref{fig:zooms} the solutions computed by wCEL0 Algorithm \ref{alg:wCEL0} in comparison with the ones obtained by CEL0 \cite{CEL02015} by Deep-STORM \cite{DeepStorm2018}, a deep-learning based model for super-resolution microscopy whose COLAB notebook\footnote{https://github.com/EliasNehme/Deep-STORM} was used to generate an ad-hoc training data using the parameters above. For a quantitative assessment of localisation precision, we compute for each reconstruction the average (over frames) Jaccard index $J_\delta\in[0,1]$ which is the ratio between correct detections (CD) up to some tolerance $\delta>0$ and the sum of CD, false negatives (FN) and false positives (FP). We test three different values of $\delta\in\left\{0,2,4\right\}$  corresponding to a tolerance of 0, $50$ and $100$ nm, respectively. Our results are reported in Table \ref{table:Jacc}, which contains also the values of CD, FN and FP computed for each method for the case $\delta=4$. We observe that the Jaccard values for wCEL0 are significantly better than the ones computed for both CEL0 and Deep-STORM. We observe that, while in terms of CD Deep-STORM outperforms the other methods, its reconstruction shows a large number of FP, as it can be observed in the close-ups in Figure \ref{fig:close-ups}. To solve this drawback (which would of course improve also the performance of CEL0 and wCEL0), post-processing techniques can be used.

\begin{table}
\centering
\small
\begin{tabular}{l|l|l|l||l|l|l|}
\cline{2-7}
                                & $J_0$ & $J_2$ & $J_4$ & CD  & FN & FP   \\ \hline
\multicolumn{1}{|l|}{CEL0}      & 0.042   & 0.467   & 0.552   & 121 & 96 & \textbf{3}    \\ \hline
\multicolumn{1}{|l|}{wCEL0}     & \textbf{0.057}   & \textbf{0.552}   & \textbf{0.659}   & 151 & 67 & 14   \\ \hline
\multicolumn{1}{|l|}{Deep-STORM} & 0.025   & 0.037   & 0.038   & \textbf{217} & \textbf{1}  & 8157 \\ \hline
\end{tabular}
\caption{Jaccard index for $\delta\in\left\{0,2,4\right\}$ and CD, FN, FP for $\delta=4$ computed as mean over the frames.}
\label{table:Jacc}
\end{table}

\vspace{-0.6cm}

\section{CONCLUSIONS}
\label{sec:conclusion}

We proposed a weighted $\ell_2-\ell_0$ model for sparse super-resolution of high-density SMLM data suited to model the presence of signal-dependent noise. To make the model tractable, we follow \cite{CEL02015} and consider its continuous exact relaxation defined in terms of a weighted-CEL0 penalty depending both on model parameters and observed data. The numerical solution of the weighted problem is challenging due to the presence of model and algorithmic parameters which are not trivial to compute. To overcome these issues, we detail suitable estimation strategies allowing to solve the problem efficiently via IRL1 algorithm. Our numerical results show improvements in molecule localisation in comparison with standard CEL0 and Deep-STORM approaches. 

Future research should address the case of general data fidelities, in order to deal directly with the case of non-symmetric terms, such as the Kullback-Leibler fidelity.

\small{\section{Compliance with ethical standards}
This work was conducted using biological data from the SMLM ISBI 2013 dataset. Ethical approval was not required as confirmed by the license attached with the open access data.}

\bibliographystyle{IEEEbib}
\bibliography{refs.bib}

\begin{thebibliography}{10}

\bibitem{Sage2015}
D.~Sage, H.~Kirshner, T.~Pengo, N.~Stuurman, J.~Min, S.~Manley, and M.~Unser,
\newblock ``Quantitative evaluation of software packages for single-molecule
  localization microscopy,''
\newblock {\em Nature methods, 12}, 2015.

\bibitem{Gazagnes2017}
S.~{Gazagnes}, E.~{Soubies}, and L.~{Blanc-Féraud},
\newblock ``High density molecule localization for super-resolution microscopy
  using {CEL0} based sparse approximation,''
\newblock in {\em IEEE ISBI 2017}, 2017.

\bibitem{CEL02015}
E.~Soubies, L.~Blanc-Féraud, and G.~Aubert,
\newblock ``A continuous exact $\ell^0$ penalty ({CEL0}) for least squares
  regularized problem,''
\newblock {\em SIAM Journal on Imaging Sciences}, vol. 8, no. 3, 2015.

\bibitem{DeepStorm2018}
E.~Nehme, L.~E. Weiss, T.~Michaeli, and Y.~Shechtman,
\newblock ``Deep-{STORM}: super-resolution single-molecule microscopy by deep
  learning,''
\newblock {\em Optica}, vol. 5, no. 4, Apr 2018.

\bibitem{Sawatzky2011}
A.~Sawatzky,
\newblock {\em (Nonlocal) Total Variation in Medical Imaging},
\newblock Ph.D. thesis, 2011,
\newblock University of M\"{unster}.

\bibitem{Carlsson2019}
M.~Carlsson,
\newblock ``On convex envelopes and regularization of non-convex functionals
  without moving global minima,''
\newblock {\em Journal of Optimization Theory and Applications}, vol. 183, no.
  1, 2019.

\bibitem{Nikolova2013}
M.~Nikolova,
\newblock ``Description of the minimizers of least squares regularized with
  $\ell_0$ -norm. uniqueness of the global minimizer,''
\newblock {\em SIAM Journal on Imaging Sciences}, vol. 6, no. 2, 2013.

\bibitem{Ochs2015}
P.~Ochs, A.~Dosovitskiy, T.~Brox, and T.~Pock,
\newblock ``On iteratively reweighted algorithms for nonsmooth nonconvex
  optimization in computer vision,''
\newblock {\em SIAM Journal on Imaging Sciences}, vol. 8, no. 1, 2015.

\bibitem{TesiMarta2020}
M.~Lazzaretti,
\newblock ``Continuous relaxation of sparse $\ell_0$ optimisation problems in
  fluorescence microscopy with {P}oisson data,''
\newblock M.S. thesis, 2020,
\newblock Universit\`a degli Studi di Genova.

\bibitem{GFISTA2019}
L.~Calatroni and A.~Chambolle,
\newblock ``Backtracking strategies for accelerated descent methods with smooth
  composite objectives,''
\newblock {\em SIAM Journal on Optimization}, vol. 29, no. 3, pp. 1772--1798,
  2019.

\end{thebibliography}

\end{document}